\documentclass[pre,aps, psfig]{revtex4}
\usepackage{epsfig}
\newcommand{\be}{\begin{equation}}
\newcommand{\ee}{\end{equation}}
\newcommand{\bn}{\begin{eqnarray}}
\newcommand{\en}{\end{eqnarray}}
\newcommand{\bd}{\begin{displaymath}}
\newcommand{\ed}{\end{displaymath}}
\newcommand{\bnn}{\begin{eqnarray*}}
\newcommand{\enn}{\end{eqnarray*}}
\begin{document}
\title{Activity-dependent self-wiring is a basis of structural plastisity in neural networks }
\author{Fail M. Gafarov}
\address{Department of Theoretical Physics, \\ Tatar State  University of Humanity and Pedagogic, \\
420021 Kazan, Tatarstan Street, 2 Russia \\
\email*{e-mails: fgafarov@yandex.ru, fail@kazan-spu.ru}}
\begin{abstract}
Dynamical wiring and rewiring in neural networks are carried out by activity-dependent growth and retraction
of axons and dendrites, guided by gudance molecules, released by target cells. Experience-dependent
structural changes in cortical microcurcuts lead to changes in activity, i.e. to changes in information encoded.
Specific pattens of external stimulation can lead to creation of new synaptical connections between neurons.
Calcium influxes controlled by neuronal activity regulates processes of neurotrophic factors release by neurons,
growth cones movement and synapse differentiation in developing neural system, therefore activity-dependent
self-wiring can serve as a basis of structural plasticity in cortical networks and can be considered as a form of learning.
\end{abstract}
 \maketitle
\section{Introduction}
The neural system is a complex self-wiring system, which consists of
a huge number of  interconnected individual cells (neurons).
Two main types of signaling exist in a neural system: short-range
synaptic signaling between neurons, provided by neurotransmitters,
which acts on postsynaptic neuron's state and long(short)-range
signaling by chemicals that acts on neuron's geometrical properties
(position of cell, dendrites, axons) \cite{Goldberg}.
Each function of the mature neural system depends on the actions of distinct  neuronal circuits
and therefore proper functioning of the neural net depends on
correctness of axonal projections and interneuronal connections \cite{CatalanoShatz}.

In  {\it in vitro} experiments it is shown that neurons self-organize into homogeneous or clustered networks
and correlations between neurons activity emerges \cite{SegevJacob}. Different models had been proposed
for investigation of connection structure emergence between initially disconnected neurons \cite{Jia,Ooyen}.

It is currently accepted that cortical maps are dynamic constructs that are remodelled in response to external input.
Two types of plasticity in neural system is known: synaptical plasticity and structural plasticity.
Synaptical plasticity involves activity-dependent weight changes between previously connected neurons \cite{Kempter}.
Structural plasticity includes remodelling of axons and dendrites, synapse formation and elimination \cite{ChklovskiiMelSvoboda}.
Neuronal activity controls metabolic process in neurons through calcium influx through voltage dependent calcium channels.
Calcium transients plays a cental role in axon guidance, neuron differentiation and synaptical plasticity (LTP,LTD) \cite{Ming,Uesaka}.

In this paper, using the framework presented in \cite{Gafarov} is demonstrated that activity-dependent self-wiring could provide
sufficient basis for structural plasticity in the developing neural system. The model proposed in \cite{Gafarov} is based on the following
experimental data (i)-(iii) and assumption (iv)-(v):
(i) Development of neuronal connectivity dependents on the neurons
activity \cite{ZhangPoo} ;
(ii) Direction of motion of the growth cones is controlled by diffusible chemicals - axon guidance molecules
(AGM) \cite{Goldberg};
(iii) Axon's growth rate dependent on the neuron's activity \cite{Ming};
(iv) Depolarization causes neurons to release axon guidance molecules \cite{Bakowec};
(v)  Type of neuronal connectivity is determined postsynaptically during synaptogenesis \cite{ZhangPoo,Spitzer}.
\section{The model and numerical results}
Axon growth is a complex process in which a growth cone, located at the tip of growing neurite travels through surrounding media,
 controlled by chemical released by targets.
The growth cone is able to detect gradients of AGM as small as a difference of a single molecule across its structure \cite{Goldberg}.
In the model suggested we assumed that, if travelling growth cone's soma is at inactive state (not depolarization and calcium influx),
is guided only by AGM, which causes the local calcium influx and the growth cone changes the direction of motion.
When the cell generates an action potential, the depolarization of its membrane leads to opening of the growth
 cone's voltage-gated calcium channels and inhibition of its growth rate \cite{Ming}.
Therefore the rate of axon's growth depends on AGM concentration gradient $\nabla C({\bf g}_k^n,t)$ at the growth cone's position \cite{Goldberg}
and axon's firing rate $\nu(t)$ \cite{Ming}, and growth cone's equation of motion can be written in the form
\begin{equation}
\frac{d {\bf g}_k^n}{dt}=\lambda F(\nabla C({\bf g}_k^n,t),\nu(t)), \label{f4}
\end{equation}
where $\lambda$ is a coefficient describing axon's sensitivity and motility.
Here we used the most simple case of the function describing axon outgrowth
\begin{equation}
\frac{d {\bf g}_k^n}{dt}=\frac{\lambda}{1+e^{-b\nu(t)}} \nabla C({\bf g}_k^n,t) \label{f4}
\end{equation}

In \cite{Gafarov} was investigated self-wiring between binary
neurons. Here we used Spike-Response Model (SRM) \cite{Gerstner}
for description of neuron's activity. Following \cite{Gerstner,
GBook} the membrane potential $u_j(t)$ of $j$-th neuron at time
$t$ is defined as \be u_i(t)=\eta(t-\hat t_j)+\sum_j w_{ij} \sum_f
\epsilon_{0}(t-t_j^{(f)})+\int_0^\infty k(s)I_i^{ext}(t-s)ds. \ee
The response kernels  $\eta$, $\epsilon_0$, $k$ that describe the
effects of spike generation, spike reception and external input on
the membrane potential are described by the following functions
\cite{GBook}: \be \eta(t-\hat t)=-v-\eta_0 \exp \left
(\frac{t-\hat t}{\tau_{refr}}\right)\Theta(t-\hat t) \ee \be
\epsilon_0(t-t_i)=\frac{1}{1-(\tau_s/\tau_m)}\left [ \exp
\left(-\frac{t-t_i}{\tau_m} \right)- \exp
\left(-\frac{t-t_i}{\tau_s} \right)\right] \Theta(t-t_i) \ee \be
k(s)=\exp\left( -\frac{s}{\tau_e}\right)\Theta(s) \ee If $u_j(t)$
crosses from below $(\frac{du}{dt}>0)$ a threshold $\theta$ at a
moment $t_j^{(f)}$ then a spike is generated.

Release of some neurotrophic factors  can be triggered by external stimulation and neuron's electrical activity \cite{Bakowec}.
In \cite{Gafarov} was proposed a hypothesis that neurons release AGM at firing time.
The activity dependent release of AGM is a key point in our model.
As far as we know in neurobiological literature there is no complete proof of the activity dependent AGM release and we consider this point as an hypothesis.
We suppose that all neurons release the unit amount of the one type AGM which causes only attraction of growth cones.
Therefore, the concentration of AGM, $c_{ij}$, released by the i-th cell at the moment $t_j$ can be found as the solution of the diffusion equation
\begin{equation}
\frac{\partial c_{ij}}{\partial{t}}=D^2 \Delta c_{ij}-k c_{ij},
\end{equation}
with the initial conditions $c_{ij}({\bf r},{\bf r}_i,
t_j)=\delta({\bf r}-{\bf r}_i)\delta(t-t_i^j)$ (point-like sources).
Here $D$ and $k$ are AGM's diffusion and degradation coefficients in the intracellular medium.
We consider here the case without boundary conditions.
The solution of this equation, describing the concentration of AGM released by a single spike of $i-th$ neuron is
\begin{equation}\label{cij}
c_{ij}({\bf r},{\bf r}_i, t, t_i^{(f)})=\frac{\Theta(t-t_i^{(f)}) }{(2 D \sqrt{\pi
(t-t_i^{(f)})})^{3}} \exp \left(-k(t-t_i^{(f)})-\frac{|{\bf r}-{\bf r}_i|^2}{4D^2 (t-t_i^{(f)})}\right). \label{diff_eq}
\end{equation}
The total concentration and gradient of concentration of AGM at the point $\mathbf{r}$ can be found by summation of concentrations
and and gradients of concentration of AGM which were released by each cell \cite{Gafarov}
\begin{equation}
C({\bf r},t)=\sum_{i=1}^{N} \sum_{f=1}^\infty c_{ij}({\bf r},{\bf
r}_i, t, t_i^{(f)}), \label{f3}
\end{equation}
\begin{equation}
\nabla C({\bf r},t)=\sum_{i=1}^{N} \sum_{f=1}^\infty \nabla c_{ij}({\bf r},{\bf
r}_i, t, t_i^{(f)}), \label{f3}
\end{equation}
where
\begin{equation}
\nabla c_{ij}({\bf r},{\bf
r}_i, t, t_i^{(f)})=\frac{\Theta(t-t_i^{(f)})({\bf r}-{\bf r_i}) }{16 D^5 \pi^{3/2}
(t-t_i^{(f)})^{5/2}} \exp \left(-k(t-t_i^{(f)})-\frac{|{\bf r}-{\bf r}_i|^2}{4D^2 (t-t_i^{(f)})}\right).
\end{equation}
If the position of some growth cone is close to the another cell's soma, i.e if $|{\bf g}_k^n-{\bf r}_i|<\varepsilon$
($\varepsilon$ can be considered as the soma's geometrical radius) then synaptic connection between these neurons will be established.

Certain neurons choose the neurotransmitter they use  and synapse type in an activity-dependent manner, and different trophic factors
are involved in this phenotypic differentiation during development.
Regulation of transmitter expression occurs in a homeostatic manner. Suppression of activity leads to an increased number of neurons
expressing excitatory transmitters and a decrease number of neurons expressing inhibitory transmitters and vice versa \cite{Spitzer}.
In the model, we supposed in framework of our approach that each neuron's different terminals (branches of the same axon) can release
different neurotransmitters and can establish different types of synaptic connections (inhibitory or excitatory).
The type of synapse can be determined by state of presynaptic or/and postsynaptic neuron.
For simplification we assumed also that the type of a synaptic connection between cells depends on the state of postsynaptic cell
at synaptogenesis process.
This assumption can be changed for special neurons, according to experimental data.
For simplicity the type of the neuronal connections between $i$-th and $n$-th neurons we describe by using
static synaptic weights $w_{n,i}$ ($w_{n,i}=1$ means  excitatory and $w_{n,i}=-1$ inhibitory connections).
According experimental data \cite{Spitzer}, in the model the type of synaptic connection established between
neurons depends on the state of postsynaptic cell at synaptogenesis moment (if $V_i(t)>V_{tr}$ then $w_{n,i}=-1$,
else $w_{n,i}=-1$).

The model gives a closed set of equations describing AGM's release and diffusion, and axons growth
 and synaptical connections establishment as well as the net's electrical activity dynamics.
The concentration of AGM in the extracellular space is controlled by neurons activity.
Growth and movement of growth cones is managed by the concentration gradients of AGM and neuronal activity.
Growth cones can make synaptic connections with other neurons and change the network's connections structure which change the network's activity.
Numerical simulation of the model were performed using the net which consist of N=8 neurons, placed at tops of a cube (Fig.1).
Initially all neurons has no synaptic connections and all axons placed near the the soma.
Different values of parameters gives different connectivity patterns between neurons, because these parameters characterize neuron's
electrical properties,and growth cone's movement speed, and AGM's acting distance, and etc.
 The figures presented here have been obtained using the following values of parameters:
 $\eta_0=0.06$, $\tau_{refr}=0.1$, $v=0.03$, $\theta=-0.02$, $\tau_s=0.01$, $\tau_m=0.2$, $\tau_e=0.1$ and $D=0.8$, $a=0.2$, $\lambda=0.00005$.
External current were taken in the form $I^{ext}_i=0.5\sin(0.2(i-4)t+i)+1$, where $i$ is the number of neuron.
After simulation start, growth cones began moving in the direction of concentration gradients (Fig. 1).
\begin{figure}
\epsfig{file=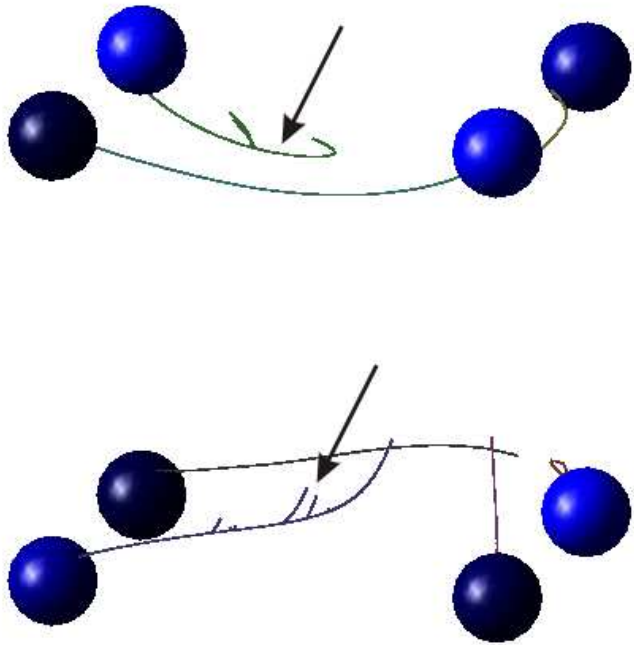,width=9cm}
\caption{Three dimensional picture of the state of the net at the simulation beginning $t=5.5$ c.
Individual neurons (whilst without neuronal connections) depicted as spheres, the firing rates depicted
by brightness (bright - hight firing rate, dark- low).
The growing axons are depicted as thin curves. One can see from this figure, how axons grow toward active
cells and that some axons began branching (new branches pointed by arrows).}
\end{figure}
Several studies show that neural activity affects individual axonal branching in vivo. Increasing of neuronal activity
leads to appearance of new branches \cite{Uesaka, Portera}.
In the model, branching of growth cone is dependent on activity (new branch added only if $V_i>10 $ $c^{-1}$). Metabolic constrains requires,
that axon cannot infinitely make new branches, to agree with it in the model a new branch is added only after some time ($0.4$ c) after previous branch addition.

When the growth cone reaches the soma of another neuron, two neurons became connected and axon's connected branch is depicted by thick curve (Fig. 2).
The synapse type (inhibitory or excitatory) is depicted by color of a curve (bright - excitatory, dark- inhibitory).
When the branch of the same axon reaches the another cell, which has already connected with it, this branch will be deleted.
\begin{figure}
\epsfig{file=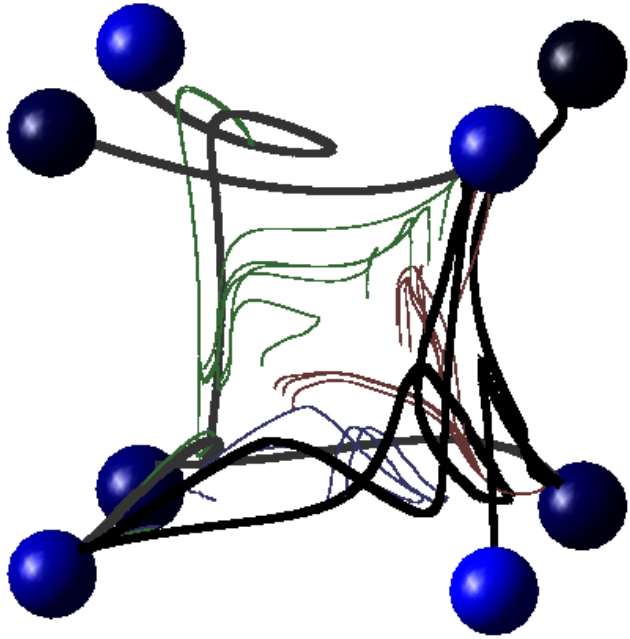,angle=0,width=9cm}
\caption{The state of the net at $t=23.3$ c. The net has weighs $w_{5,7}=1$, $w_{2,4}=1$,$w_{8,4}=-1$, $w_{3,4}=-1$, $w_{1,6}=1$, $w_{7,6}=-1$, $w_{6,7}=-1$.}
\end{figure}
Structural changes leads to changes in electrical activity of neurons, i.e to changes in the time dependence of membrane potentials $u_j(t)$ (Fig. 3).
\begin{figure}
\epsfig{file=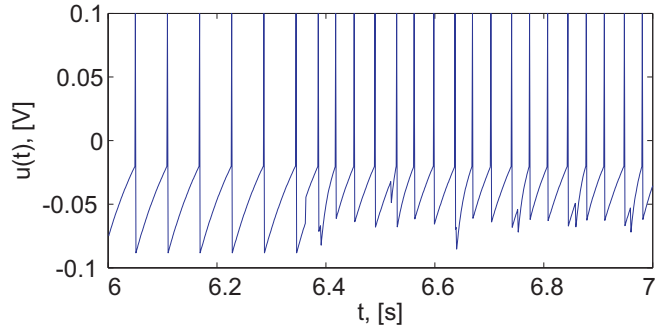,width=9cm}
\caption{The time dependence of membrane potential of 5-nd neuron. One can see postsynaptic potentials, and firing rate
increasing after $t=6.35$ c because of new synaptic connection between 7-nd and 5-nd neurons.}
\end{figure}
\section{Conclusions}
Neural activity plays a central role in experience-dependent rewiring of cortical microcircuits.
Activity-dependent structural plasticity can be considered as a special case of synaptical plasticity in fully connected network.
Both of these types of plasticity are dependent on activity pattern of neurons. But in contrast to synaptical plasticity structural
plasticity imply changes in connection map.
Real neural net are sparsely connected and in no circumstances we cannot  consider them as fully connected network.
Activity-dependent self-wiring involves establishment of new synaptic connections between  previously unconnected cells, and this
conception is important for investigation of learning in real networks.
In the model developed here, finding of appropriate partnership between pre- and postsynaptic neuron is controlled by activity of neurons,
therefore different patterns of external input $I^{ext}_i$ through regulation of neuronal activity will lead to functionally distinct circuits,
and changes in connections structure can lead to changes in activity (Fig. 3) of the wholly network.
For the further theoretical and computational investigations of structural plasticity in neural networks the model presented here, can be
sophisticated on the basis of new experimental findings, for example:
\begin{itemize}
\item Cells can release also repellant in activity-dependent or activity-independent manner \cite{Goldberg}, therefore a
model where at inactive state cells release a repellant, at active state - an attractant, or vice versa, can be considered.
\item A set of cells releasing different types of AGM, and different types of growth cones regulated by different AGM can be considered \cite{Goldberg}.
\item Depending on the cell's level of activity, a growth cone can be repelled or attracted by the same AGM \cite{Ming}.
\item Growth cones themselves can release chemicals and influence other growth cones movement \cite{HenOoyen}.
\item A real neurons has also dendrites. For simplification of model we did not consider them, and supposed that axons connect
directly to a soma. Incorporation into model also guidance of dendrites \cite{Scott} by different chemicals, can cause discovery
of a new interesting properties of structural plasticity.
\item Hebbian learning can be incorporated \cite{Kempter}.
\end{itemize}
We believe that the model developed here  may help in investigations of fundamental problems in neural networks self-organization
{\it in vivo} and {\it in vitro} \cite{SegevJacob}. Specially, this model can be used in construction of novel biosensors and hybrid neural-computer systems.

\end{document}